# Spin selectivity through chiral polyalanine monolayers on semiconductors

Tianhan Liu[1], Xiaolei Wang[2,4], Hailong Wang[2], Gang Shi[3], Fan Gao[3], Honglei Feng[3], Haoyun Deng[1], Longqian Hu[1], Eric Lochner[1], Pedro Schlottmann[1], Stephan von Molnár[1], Yongqing Li[3], Jianhua Zhao[2]*, and Peng Xiong[1]*

[1]*Department of Physics, Florida State University, Tallahassee, Florida 32306, USA*

[2]*State Key Laboratory of Superlattices and Microstructures, Institute of Semiconductors, Chinese Academy of Sciences, Beijing, 100083, China*

[3]*Beijing National Laboratory for Condensed Matter Physics, Institute of Physics, Chinese Academy of Sciences, Beijing, 100190, China*

[4]*College of Applied Sciences, Beijing University of Technology, Beijing 100124, China*

*Email: jhzhao@red.semi.ac.cn, pxiong@fsu.edu

**Electrical generation of polarized spins in *nonmagnetic* materials is of great interest for the underlying physics and device potential. One such mechanism is chirality-induced spin selectivity (CISS), with which structural chirality leads to different electric conductivities for electrons of opposite spins. The resulting effect of spin filtering has been reported for a number of chiral molecules. However, the microscopic mechanism and manifestation of CISS in practical device structures remain controversial; in particular, the Onsager relation is understood to preclude linear-response detection of CISS by a ferromagnet. Here, we report direct evidence of CISS in two-terminal devices of chiral molecules on the magnetic semiconductor (Ga,Mn)As: In vertical heterojunctions of (Ga,Mn)As/AHPA-L molecules/Au, we observed characteristic linear- and nonlinear-response magnetoconductance, which directly verifies spin filtering by the AHPA-L molecules and spin detection by the (Ga,Mn)As. The results constitute definitive signature of CISS-induced spin valve effect, a core spintronic functionality, in apparent violation of the Onsager reciprocity. The results present a promising route to semiconductor spintronics free of any magnetic material.**

Recently, there has been growing interest in electronic methods of producing spin polarization in semiconductors (SCs) *without* using any magnetic materials. One pathway is via spin-orbit interactions (SOI) with which electron charge motion in a specific direction leads to spin polarization in an orthogonal orientation. Examples include spin Hall effect in III-V SCs[1–3] and spin-momentum locking in spin-helical surface states in 3D topological insulators[4]. Another scheme utilizes charge motion through materials exhibiting structural chirality in real space. The effect, termed chirality-induced spin selectivity (CISS)[5,6], has been reported in a variety of chiral molecules including dsDNA[7,8], polypeptides[9,10] and helicenes[11]. In contrast to solid state materials, organic molecules exhibit rich variety of structures, which can be readily tailored to realize wide-ranging electronic properties and functionalities favorable for spintronics[12]. The experiments on CISS generally involve a self-assembled monolayer (SAM) of chiral molecules on a non-magnetic noble metal. Spin filtering or spin selective transport of the electrons from the noble metal through the chiral SAM is evidenced in photo-electric[7], scanning conductance microscopy[8], fluorescence microscopy[13], and voltammetry[14] measurements. Spin polarization as high as 60% was measured at room temperature[7]. Besides chiral molecules, CISS was also predicted and/or observed in carbon nanotubes decorated with chiral molecules[15,16,17] and 2D chiral hybrid perovskites[18].

Theoretical studies of CISS have focused primarily on its microscopic origin, especially the relevance of SOI and molecular level structural details of the chiral molecules[19–23].

In contrast, modeling of the manifestation of CISS in practically important transport devices has been scarce and inconsistent. Two recent such studies[24,25] have produced contradictory results by modeling the magnetoconductance (MC) of chiral molecule junctions and taking into account effects from the metal electrodes and their contacts with the molecules.

Experimentally, the device potential of CISS was demonstrated in a type of proof-of-concept memory device[9,10], where the α-helical polyalanine was shown to be able to facilitate the magnetization reversal of a magnetic layer without any electron transfer. However, CISS devices involving *active* electron transport through the chiral molecules, e.g., those resembling a magnetic tunnel junction or a spin valve, have not been demonstrated. Conceptually, such a device would consist of a normal metal (NM) and a ferromagnet sandwiching a chiral molecule SAM, and CISS of the molecules would manifest in a MC corresponding to the magnetization reversal in the FM. So far, such a MC has been seen in the setup of conductance atomic force microscopy (cAFM)[8], which is obviously not amenable to practical applications. Moreover, the conductance measurement depends to a large extent on the contact of the cAFM tip and a molecule, which leads to significant fluctuations and the necessity of replying on statistical averages of great number of measurements. For the practical rendition of *planar* junctions, a critical obstacle is the well-known one in the field of molecular electronics: A SAM cannot serve as an insulating barrier over practical device length scales, and any direct contact of the two metal electrodes through defects in the SAM fully shorts out the device. To circumvent this problem, an oxide layer was added between the two metal layers[9,11,26]. Magnetoresistance (MR) was observed in cross-stripe planar junctions of Ni/Al$_2$O$_3$/chiral molecules/Au with oligopeptides[26] and helicenes[11] molecules. However, the origin of the observed MR remains unclear. The insertion of the oxide barrier brings about additional complications. The authors attributed the CISS in these devices to the chirality of the Al$_2$O$_3$ deposited onto the chiral molecules[26], which is in apparent contradiction to the microscopic theories depending on internal structures of the chiral molecules[19–23].

Here, we report direct evidence for CISS of chiral molecules assembled on a semiconductor by measuring the MC of *vertical planar* junctions of (Ga,Mn)As/ α-helix L-polyalanine (AHPA-L) molecules/Au. The experiments were made possible by our ability to create high-quality SAMs and their micro/nano-patterns on GaAs[27,28]. By replacing one of the metal electrodes with a doped semiconductor, the Schottky barrier at the metal/SC direct contact effectively mitigates electrical shorts through defects in the SAM, as demonstrated in molecular junctions on $p^+$-GaAs[29]. The utilization of this device scheme makes the electrical conductance through molecular SAM experimentally discernible from the overall junction conductance, and makes possible the observation of MC distinctly associated with CISS in the two-terminal device. The device structure also facilitates reliable examination of the bias dependences of the MC, which reveal *both a pronounced nonlinear-response and a nontrivial linear-response component in the MC*, in apparent violation of the Onsager reciprocity[30]. The experiment directly verifies the veracity of CISS in realizing a core spintronics functionality, spin filtering of the electrons injected from Au through the chiral molecules as detected by the ferromagnetic (Ga,Mn)As. The realization of CISS in a SC-based two-terminal device points to the potential of CISS in semiconductor spintronics for spin injection and detection without using any magnetic materials.

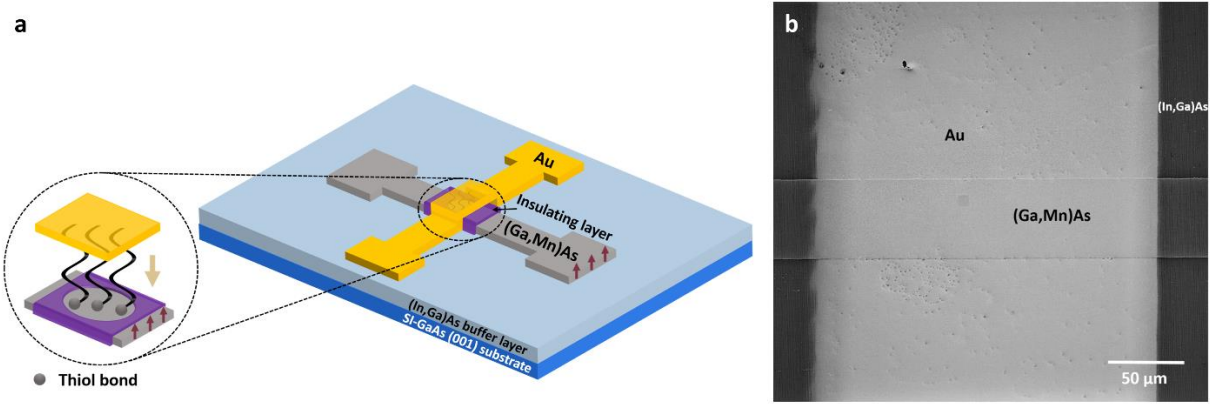

**Fig. 1 | (Ga,Mn)As/chiral molecules/Au junction. a**, Schematics of the device structure. The close-up image depicts the internal structure of a (Ga,Mn)As/AHPA-L molecules/Au vertical junction. The garnet arrows indicate the perpendicular magnetic anisotropy in the (Ga,Mn)As. The gold arrow indicates the spin polarization direction of the electrons through chiral molecules, which is always along the helical axis of the molecule. AHPA-L molecules are assembled on (Ga,Mn)As with the thiol bond. The vertical junction is defined by a small opening in the insulating layer fabricated by electron beam lithography. The Cr/Au electrode is evaporated on top through a shadow mask. **b**, Scanning electron microscopy image of a 5×5 μm$^2$ junction (the dark square in the central region).

Fig. 1a,b show a schematic diagram and scanning electron microscopy (SEM) micrograph, respectively, of a vertical junction of (Ga,Mn)As/AHPA-L molecules/Au. Here, the α-helix has 3.6 amino acids per turn of the helix and the distance between each turn is 0.54 nm, thus the length of the AHPA-L is 5.25 nm. The cysteine at the N-terminus contains thiol, which can form covalent bonds with Ga and As[27], facilitating the formation of SAM on (Ga,Mn)As. The AHPA-L molecules were assembled on MBE-grown epitaxial (Ga,Mn)As via solution assembly, and the resulting molecular layer was measured via ellipsometry to have a thickness of 3.3 nm on GaAs (see **Experimental Section**). The result indicates that the molecules form a monolayer with the molecules tilting at an angle of 51° with respect to the normal, larger than the reported value of 40° on Au[10].

The (Ga,Mn)As was grown by low-temperature MBE on an (In,Ga)As buffer layer; the resulting tensile strain leads to perpendicular magnetic anisotropy (PMA)[31]. The as-grown (Ga,Mn)As thin film used in this study has a Curie temperature of 140 K and coercive field of 460 Oe (details in the Supplementary Information S1). The actual coercive fields of the (Ga,Mn)As in different devices varied from 180 to 460 Oe depending on the specific annealing conditions they were subject to. The devices were fabricated via a process consisting of photo- and electron beam lithography, Argon ion milling, AHPA-L SAM assembly, and top electrode evaporation. In Fig. 1b, a close-up SEM image shows a 5×5 μm² junction in a fabricated device.

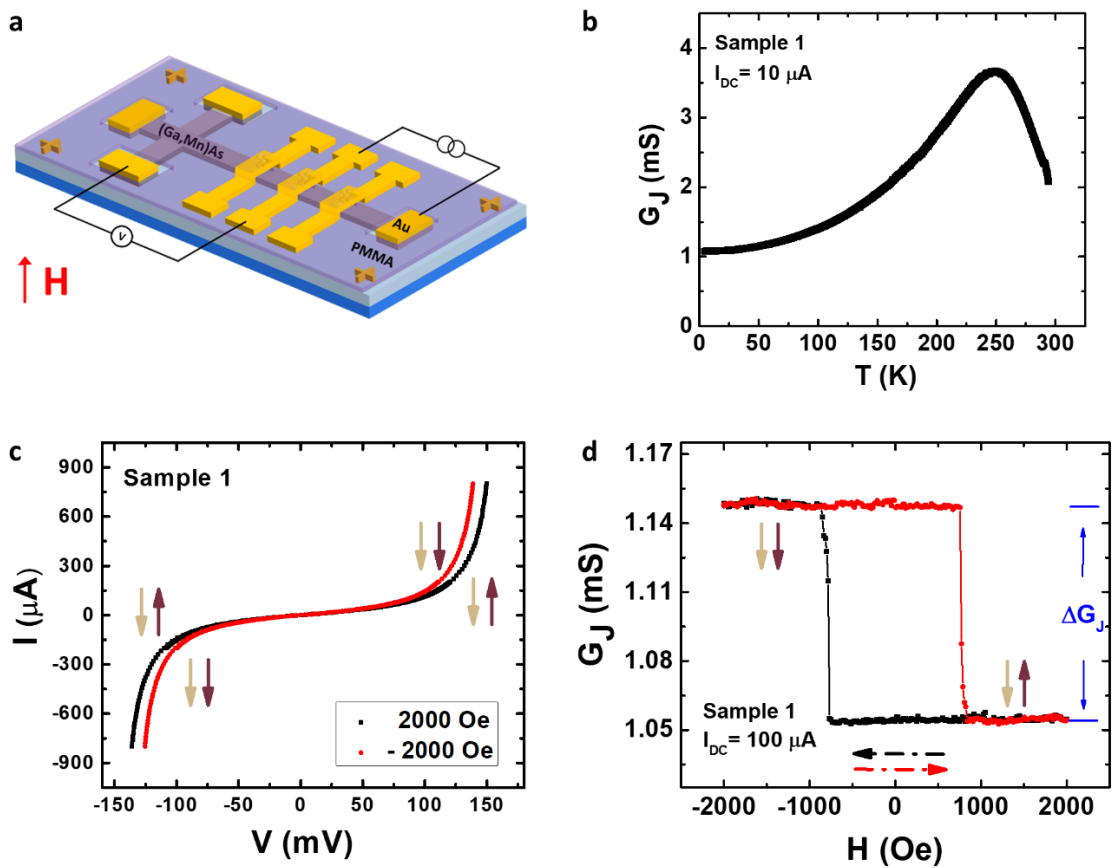

**Fig. 2 | Electrical measurements of (Ga,Mn)As/AHPA-L molecule/Au junction. a**, A schematic diagram of the experimental setup for junction measurements. **b**, Low-bias junction conductance as a function of temperature in zero applied magnetic field. **c**, I-V characteristics of the junction in perpendicular magnetic fields of ±2000 Oe (greater than the coercive field of the (Ga,Mn)As). **d**, Junction conductance versus perpendicular magnetic field measured at a temperature of 4.2 K and a DC bias of 100 μA. Both the I-V and MC measurements show two distinct conductance states depending on the direction of the (Ga,Mn)As magnetization as indicated by the garnet arrow. The gold arrow indicates the direction of the electron spin polarization, which is independent of the magnetic field. Measurements in **c**, **d** were performed at 4.2 K.

A schematic diagram of the magneto-electrical measurement setup for the (Ga,Mn)As/AHPA-L molecule/Au junctions is shown in Fig. 2a. The specific junction resistance (RA product) values are on the order of $10^2$ k$\Omega\cdot\mu m^2$, which are about an order of magnitude or higher than the typical values for control junctions without the AHPA-L molecule SAM. Fig. 2b,c show a set of results from a 10×10 $\mu m^2$ junction in sample 1. The zero-field low-bias junction conductance decreases with decreasing temperature at low temperatures and eventually saturates (Fig. 2b). The general insulating behavior of the junction conductance and the high specific junction conductance at low temperature indicates significant coverage of the AHPA-L molecules on (Ga,Mn)As. It is also consistent with the quantitative estimation with the junction resistances with/without AHPA-L molecules. (See Supplementary Information Note 1 for details). The field-dependent I-V characteristics and perpendicular field MC of the junction were measured at low temperatures, and the results at 4.2 K are shown in Fig. 2c and Fig. 2d respectively. The I-V curves show strong nonlinear behavior, consistent with an asymmetric barrier which results in higher order contributions to the junction conductance[32]. A clear split was observed for the I-V curve when perpendicular magnetic fields of opposite polarities, ±2000 Oe, were applied. Since the applied fields were much above the coercive field of the (Ga,Mn)As, the two distinct conductance states are clearly associated with the reversed magnetization of the (Ga,Mn)As. This is evidenced directly in the MC measurement shown in Fig. 2d.

The MC was obtained by measuring the voltage across the junction at a fixed bias current (100 µA in this case) while sweeping the perpendicular magnetic field. Fig. 2d shows a typical MC response in the form of sharp changes of the junction conductance coinciding with the coercive fields of the (Ga,Mn)As. The sharp conductance jumps are a result of the PMA in the strained (Ga,Mn)As. The dashed arrows refer to the directions of the field sweep for the red and black MC curves. We attribute the distinct two-state MC to a direct consequence of the CISS of AHPA-L molecules: The unpolarized electrons from the Au electrode attain a spin polarization as they transport through the chiral molecules. The sign of the spin polarization depends only on the helicity of the molecules and is independent of the external magnetic field (gold arrow)[21,23]. The (Ga,Mn)As, with high intrinsic spin polarization in the ferromagnetic state[33], acts as a spin analyzer. As a result, the junction conductance changes as the magnetization of the (Ga,Mn)As is flipped. Here the percentage change of the junction conductance is about 9%; however, the junction conductance likely consists of parallel contributions from transport through the chiral molecules and defects in the SAM ((Ga,Mn)As/Au direct contact), hence the absolute change of the junction conductance, $\Delta G_J$, rather than the percentage change, is a more accurate measure of the CISS effect.

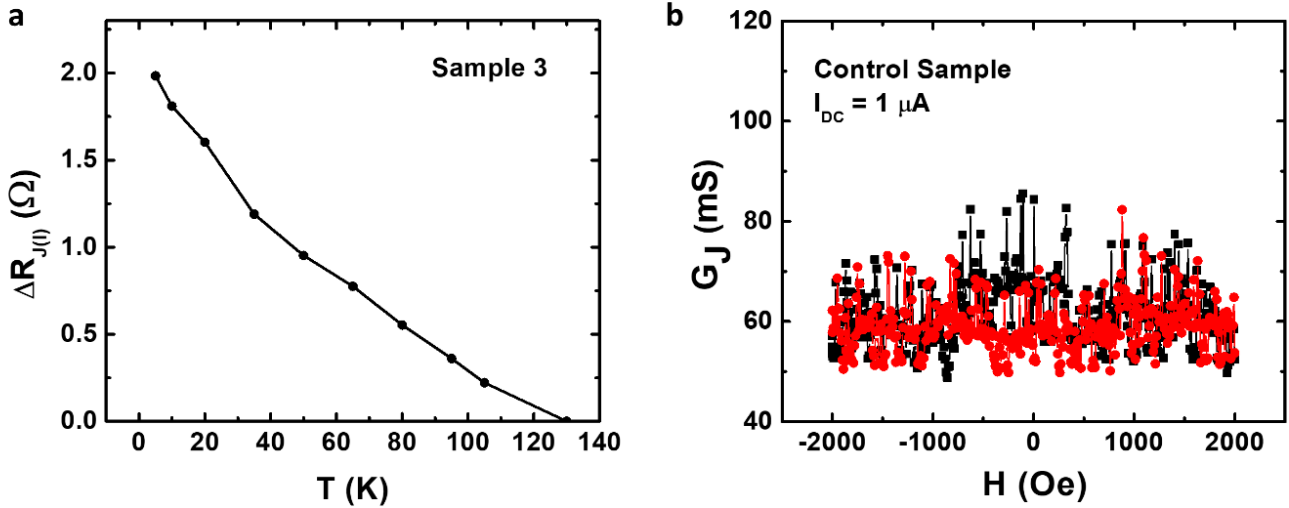

**Fig. 3 | Temperature dependence of junction magnetoresistance and control junction MC. a**, $\Delta R_{J(I)}$ as a function of temperature. The data points were calculated based on $\Delta R_{J(I)} = \frac{V_\uparrow - V_\downarrow}{I}$. Measurements were taken at 5 K, 10 K, 20 K, 35 K, 50 K, 65 K, 80 K, 95 K, 105 K, and 130 K. **b**, junction conductance versus perpendicular magnetic field of a control junction without chiral molecule SAM. No measurable MC is present.

Fig. 3a shows a representative temperature dependence of the junction MR. As expected, $\Delta R_{J(I)}$ decreases with increasing temperature and vanishes at the Curie temperature of the (Ga,Mn)As. Qualitatively, the decrease of $\Delta R_{J(I)}$ with increasing $T$ is more rapid than that of the magnetization, and is consistent with the faster decreasing spin polarization in (Ga,Mn)As.

In order to establish a definitive connection between the observed MC and electron transport through the chiral molecules, control devices were fabricated with the same process omitting the assembly of AHPA-L molecules on the junctions. The same set of measurements were performed in several different control samples, showing qualitatively similar results. The absence of distinct conductance states is shown directly in Fig. 3b for a control junction; the junction conductance has no measurable changes at the coercive fields and the value is consistently an order of magnitude or larger than those of junctions with molecules. These results unambiguously point to the molecules as the origin of the observed spin-dependent MC.

It is also worth noting that although the junctions contain a layer of organic molecules as a critical component, their electrical characteristics are remarkably stable under ambient conditions. The measurements performed on a device after being stored in a desiccator at room temperature for four months yielded essentially the same results as those from measurements right after its fabrication (within a day). (See Fig. S2 in Supplementary Information).

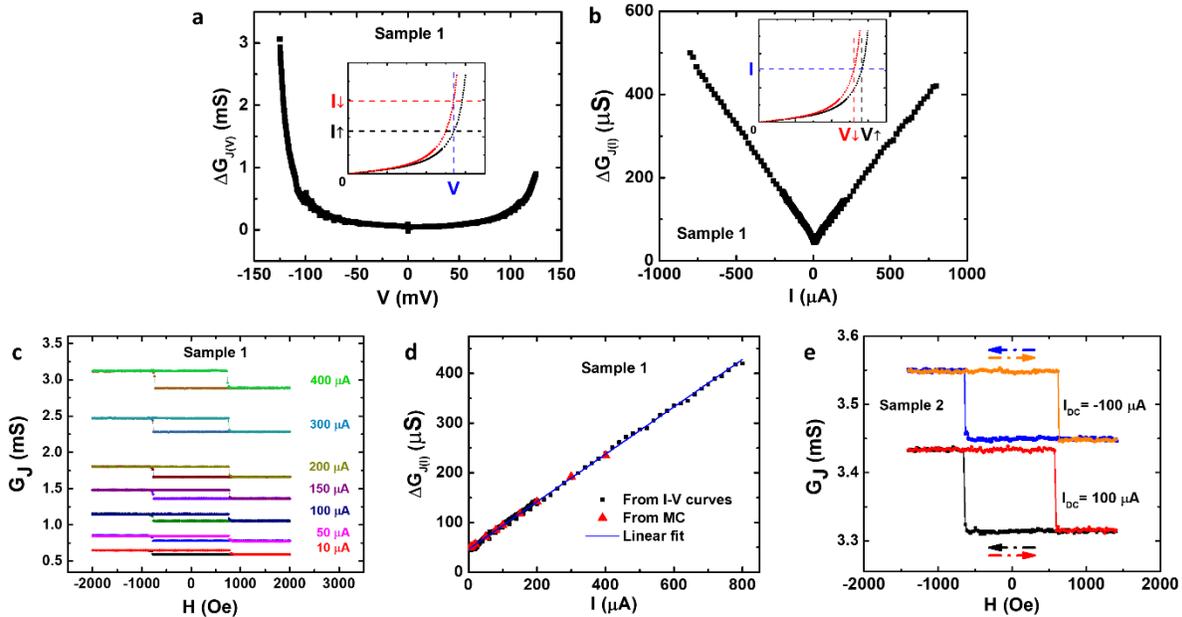

**Fig. 4 | Bias dependences of junction magnetoconductance.** $\Delta G_J$ as a function of bias voltage (**a**) and bias current (**b**). The black squares in **a** and **b** are determined from the I-V curves in Fig. 2c, with fixed voltage and current, respectively. The respective insets illustrate how the data are extracted. **c**, Representative MC curves measured at different bias currents. **d**, The black squares are from the positive currents in **b**. The red triangles are data from MC measurements at different bias currents (some are shown in **c**). The blue line is a linear fit to the black squares. **e**, MC curves for sample 2 at DC biases of 100 μA and -100 μA.

We now turn to the origin of the MC and its implications on the theoretical models. One most pertinent unsettled issue in the field is whether the CISS could lead to measurable spin valve effect in a two-terminal device. Yang *et al.* [24] note that a spin-flip electron reflection process is inherent in the CISS transport in order to satisfy the Onsager relation. This implies a vanishing MC in the linear response regime in a two-terminal junction of nonmagnetic-metal/chiral molecule/ferromagnet; rather, a four-terminal nonlocal scheme is required for the MC to materialize[24]. However, Dalum and Hedegård[25] recently examined the spin-dependent electron transport of a similar system and reached the opposite conclusion. They argue that SOI in the chiral molecules breaks the Onsager reciprocity and a new equilibrium state emerges with CISS-induced spin accumulation in the nonmagnetic lead. A nontrivial MC results from the emergent equilibrium state, which resembles that in a magnetic tunnel junction with two magnetic electrodes. In our devices, by measuring the bias dependences if the MC, we observed *both a pronounced nonlinear response MC and a nontrivial linear-response MC*, as shown in Fig. 4.

Analytically, Dalum and Hedegård calculated the electric current through the junction[25], $I_\uparrow$ and $I_\downarrow$ (the arrows indicate the directions of the magnetization in the (Ga,Mn)As lead), based on the Landauer-Buttiker formalism. Upon magnetization reversal in the magnetic lead, $I_\uparrow$ and $I_\downarrow$ are different due to the difference in the induced magnetization from the spin accumulation in the nonmagnetic lead. Using the convention in Ref. 25:

$$I_{\uparrow\downarrow} = \int T_{LR}^0 (1 \pm A\vec{m} \cdot \vec{a}_{\uparrow\downarrow})(n_F(E - \mu_L) - n_F(E - \mu_R)) \frac{dE}{2\pi}, \quad (1)$$

where $T_{LR}^0$ is the transmission function between the magnetic and nonmagnetic leads satisfying the Onsager relation, $A$ is a function related to SOI, $\vec{m}$ is the unit vector along the magnetic moment direction in the magnetic lead, and $n_F$ is the Fermi-Dirac distribution. (For details, see Supplementary Information Note 3). $\vec{a}_\uparrow$ and $\vec{a}_\downarrow$ are the induced magnetizations from spin accumulation in the nonmagnetic lead for opposite magnetizations in the magnetic lead. Because of the CISS, $\vec{a}_\uparrow \neq -\vec{a}_\downarrow$, and the transmission coefficients for the opposite magnetizations, $T_{\uparrow\downarrow} = T_{LR}^0 (1 \pm A\vec{m} \cdot \vec{a}_{\uparrow\downarrow})$, are different, with $\Delta T = T_\uparrow - T_\downarrow = T_{LR}^0 A\vec{m} \cdot (\vec{a}_\uparrow + \vec{a}_\downarrow)$. Therefore, Eq. 1 implies two distinct conductance states depending on the magnetization direction of the magnetic lead with the transition at the coercive field of the (Ga,Mn)As. This could be the origin of the nontrivial MC in our junctions.

The veracity of the model can be further tested from the dependence of the MC on the magnitude and direction of the current/voltage bias across the junction. A set of experimental results are shown in Fig. 4. The data in Fig. 4a, b (black squares) are extracted from the I-V curves in opposite saturation fields in Fig. 2c, for fixed voltage and current respectively (the ways the data are extracted are illustrated in the respective insets).

Fig. 4a shows an approximately voltage independent finite MC at low biases, which rises sharply at biases coinciding with the turn-on voltage in the junction I-V. Each bias voltage corresponds to two different current states. From Eq. 1, we can derive an explicit expression for the bias-dependent MC:

$$\Delta G_{J(V)} = \frac{I_\downarrow - I_\uparrow}{V} = \int \alpha \Delta T \frac{dE}{2\pi} + \mathbf{V} \int \beta \Delta T \frac{dE}{2\pi} + \mathbf{V^2} \int \gamma \Delta T \frac{dE}{2\pi}, \quad (2)$$

where $\mu_R - \mu_L = eV$, with $V$ being the bias voltage across the junction, and α, β, γ are energy and temperature dependent coefficients. (See Supplementary Information Note 3 for details). The finite zero-bias MC in Fig. 4a is consistent with Eq. 2. The sharp rise at high biases implies that the higher order terms in Eq. 2 should be significant; however, quantitatively, the rapid increase of $\Delta G_{J(V)}$ at the turn-on voltage is much closer to exponential than power law.

In contrast to the somewhat complex voltage dependence, $\Delta G_{J(I)}$ shows a striking linear dependence on the bias current, as shown in Fig. 4b, 4d. Here, $\Delta G_{J(I)} = I(\frac{1}{V_\downarrow} - \frac{1}{V_\uparrow})$. It is important to note that *the linear current dependence spans the entire bias range*, across the two distinct regimes in the voltage dependence. Similar voltage and current dependences are obtained in multiple junctions (another example is shown in Supplementary Information S3).

The MC at different fixed currents can also be determined from full MC sweeps, and some representative MC curves are shown in Fig. 4c. The $\Delta G_{J(I)}$ from these measurements are plotted as red triangles in Fig. 4d, which are in excellent agreement with the black squares from the positive-bias I-V's in Fig. 4b. A linear fit to the current-dependence data (blue line) yields $\Delta G_{J(I)} = \Delta G_{J(0)} + aI$, where $\Delta G_{J(0)} = 46.2 \ \mu S$. The value of $\Delta G_{J(0)}$ is consistent with the finite zero voltage bias value and constitute direct evidence that *a linear-response spin-valve type MC is present in this two-terminal device*, although the higher-order nonlinear contributions appear substantial.

Fig. 4e shows MC for another junction at DC currents of opposite polarities, ±100 μA, where +100 μA indicates the current flowing from the (Ga,Mn)As substrate to Au and vice versa. For this junction, a high-field hysteretic symmetric background is present, which has been subtracted to show $\Delta G_J$ clearly (see Supplementary Information S5 for details). Here, it is evident that the *reversal of the DC current direction does not change the junction conductance states* or, equivalently, the sign of the spin polarization. However, it does change the magnitude of the MC slightly; specifically, $\Delta G_{J,I_+} = 0.118 \ mS$ and $\Delta G_{J,I_-} = 0.101 \ mS$. This is consistent with the apparent asymmetry between positive and negative currents in Fig. 4b. The overall junction conductance also shifts by $0.124 \ mS$ upon reversing the bias current, i.e., $G_{J,I_+} - G_{J,I_-} = 0.124 \ mS$, where $G_{J,I_\pm}$ is the conductance at positive/negative DC current. The differences between $\Delta G_{J,I_+}$ and $\Delta G_{J,I_-}$, and between $G_{J,I_+}$ and $G_{J,I_-}$, are consistent with the presence of a first-order (odd) term in Eq. 2 and Eq. S4 (Supplementary Information), respectively. The bias dependences of the MC, especially the striking linear current dependence, should place significant constraints on any theoretical model of the CISS-induced spin filtering and warrant further investigation.

In summary, we have obtained direct evidence of CISS through chiral molecules assembled on a semiconductor surface. Experimentally, the CISS effect manifests in clear spin valve signals in (Ga,Mn)As/AHPA-L molecules/Au junctions, resulting from spin filtering by the AHPA-L monolayer. Nontrivial linear- and nonlinear-response CISS-induced spin valve signals are clearly identified in the two-terminal devices. The observation appears to violate the Onsager reciprocity, which should be accounted for in any viable theory for CISS and its device manifestations. With high spin filtering efficiency at room temperature[7], the realization of CISS in chiral molecules and its detection by a magnetic semiconductor present a promising nonmagnetic pathway to spin injection and detection in semiconductor spintronics devices.

## Experimental Section

**Materials and sample preparation**

The AHPA-L in the experiments was purchased from RS synthesis, LLC. It is based on α-helix L amino acids (H-CAAAA KAAAA KAAAA KAAAA KAAAA KAAAA KAAAA K-OH), where C, A, and K represent cysteine, alanine, and lysine. α-helix has a right hand-spiral conformation. The AHPA-L molecules were dissolved in pure ethanol at 1 mM concentration. The solution was kept at -18 °C for storage.

The perpendicularly magnetized (Ga,Mn)As films were grown by MBE. A 500 nm-thick (In,Ga)As buffer layer was first grown at 450 °C on semi-insulating (001) GaAs substrates. 40 nm-thick (Ga,Mn)As films with perpendicular magnetic anisotropy were later grown at substrate temperature of 270 °C. The Curie temperature as-grown varies from 20 K to 90 K depending on the Mn concentration (4% or 6%) and growth temperature. The carrier density is from $5 \times 10^{20}$ $cm^{-3}$ to $1 \times 10^{21}$ $cm^{-3}$. The Curie temperature increased from 90 K up to 144-149 K after annealing. The coercive field varies from 180 Oe to 460 Oe depending on the annealing conditions.

For ellipsometry measurements, (Ga,Mn)As samples were first soaked in ammonium polysulfide solution at 50 °C for 5 min to remove the native oxide layer on the surface.[28] They were later left in the AHPA-L solution for 24 hours for molecular self-assembly at room temperature. They were rinsed with ethanol and blown dry with nitrogen gas after the assembly.

**Fabrication process**

The junction devices were fabricated in the following steps:

a) Define the (Ga,Mn)As channel

The (Ga,Mn)As channel was first defined by photolithography. The sample was spin-coated with photoresist AZ5214E and prebaked at 110 °C for 50 s on a hotplate. It was later exposed under 350-500 nm UV light for 10 s and developed in a 1:5 solution of sodium-based AZ 351 developer diluted in DI water for about 2 min. After developing, the sample was post-baked at 120 °C for 60 s on the hotplate.

Then the electrode was etched by ion milling with an Ar ion beam produced by a 2'' Kaufmann source. The Ar flow rate was 8.8 sccm, resulting in a pressure of $1.3 \times 10^{-3}$ torr. The discharge voltage was 62.5 V, the acceleration voltage was 210 V, and the beam voltage was 500 V. The cathode current was 6.8 A and the beam current was 20 mA. Atomic force microscopy (AFM) measurements showed 73-76 nm etching depth with 7 min of milling.

b) Deposit Au contacts and alignment marks

A set of alignment marks were defined by photolithography with the same parameters as in step (a). Cr and Au. Cr (5 nm) and Au (20 nm) were then deposited by thermal evaporation, both at rate of 1 Å/s. After the evaporation, the sample was immersed in acetone overnight for lift-off, followed by rinsing with acetone and isopropanol.

c) Define junctions by electron-beam lithography (EBL)

The sample was spin-coated with 2% PMMA at 4 krpm for 30 s. It was prebaked at 180 °C for 10 min on a hotplate. The EBL was performed with an acceleration voltage of 20 kV and targeted dose of 160 μC/cm². For the small junction patterns, the spot size was 1.0, the step size was 0.01 μm and the beam current was 0.0223 nA; for the large contact patterns, the spot size was 4.0, the step size was 0.05 μm and the beam current was 0.825 nA, as measured by a Faraday cup. The sample was developed in methyl isobutyl ketone (MIBK) diluted with isopropanol (1:3) for 40 s and then in pure isopropanol for 30 s at room temperature.

d) Remove oxide layer on (Ga,Mn)As and assemble AHPA-L on the junctions

The sample was cleaned with $O_2$ plasma to remove any organic residue. It was set with medium power at 200 mtorr oxygen pressure for 1 min. The sample was then baked at 180 °C for 20 min on the hotplate to harden the PMMA. To remove the native oxide layer on the (Ga,Mn)As, the sample was etched with ion mill for 1 min with the same parameters as in step (a). It was immersed in ethanol immediately after being taken out from the ion mill chamber before the molecular assembly. Here, we chose a different method for oxide removal than the ammonium polysulfide passivation of (Ga,Mn)As used before,[27,28] as we noticed that the ammonium polysulfide tends to contaminate the surface after leaving the sample in the solution at 50 °C for 5 min.

For the assembly of AHPA-L monolayer on the (Ga,Mn)As, the sample was left in the AHPA-L solution at room temperature for 24 hours. After the assembly, the sample was rinsed with ethanol and dried with nitrogen gas.

e) Deposit top Au electrodes

A shadow mask was positioned on top of the sample by aligning the electrode patterns with the junctions under an optical microscope. For the evaporation, the same parameters as in step (b) were used in this step, except for the thickness of Au (50 nm). Also, the substrate was cooled with liquid nitrogen during the evaporation. The substrate temperature was maintained at -110 °C.

The control samples were fabricated in an identical process, except that in step (d) the sample was immersed in a pure ethanol instead of the AHPA-L molecule solution.

The samples were stored in a desiccator to minimize exposure to ambient moisture after fabrication.

**Measurements**

**Electrical measurements**

The sample was fixed on a socket with a copper base with GE vanish or photoresist, and wired by hand with silver paint and Pt wire. The sample was measured within a few days after fabrication in a Janis $^4$He cryostat and/or an Oxford $^3$He cryostat. All the measurements were performed at 4.2 K unless otherwise noted. Magnetic field perpendicular to the sample plane was applied up to 2000 Oe. The (Ga,Mn)As was first magnetized at 2000 Oe and then the magnetic field was swept at a constant rate of 400 Oe/min for measurements. DC measurements were done with Keithley 2400 as the current source and HP 3458 as the voltmeter. AC measurements were performed with EG&G 124A and/or SR2124 dual-phase analog lock-in amplifiers. The sample was later measured at increased temperatures with similar procedure.

To measure the temperature dependence, the sample was cooled from 300 K to base temperature. The resistance was measured with DC current. For every 0.1 K change in temperature, positive and negative currents were applied to the sample. The resistance was determined to be the voltage difference divided by twice of the applied current. The current reversal was necessary to eliminate the thermoelectric voltages along the circuit.

**Ellipsometry measurements**

The ellipsometry measurements were performed with an M-2000 Spectroscopic Ellipsometer. Five spots were chosen at random on two different oxide-free GaAs samples after the molecular assembly. The thickness of the monolayer was measured to be 3.3 nm for all spots.

**Acknowledgment**

We thank Kate Carnevale, Siwei Mao, Shucheng Tong, and Shengzhi Zhang for technical assistance. We acknowledge helpful discussions with Oren Ben Dor, Hanwei Gao, Zhilin Li, Yu Miao, Ron Naaman, Geoffrey Strouse, Qing-feng Sun, and David Van Winkle.

The work at FSU is supported by NSF grant DMR-1905843.

The work at Chinese Academy of Sciences is supported by National Natural Science Foundation of China (NSFC, Grant No. 11674312) at Institute of Semiconductors.

# Supplementary Information

**Spin selectivity through chiral polyalanine monolayers on semiconductors**

Tianhan Liu[1], Xiaolei Wang[2,4], Hailong Wang[2], Gang Shi[3], Fan Gao[3], Honglei Feng[3], Haoyun Deng[1], Longqian Hu[1], Eric Lochner[1], Pedro Schlottmann[1], Stephan von Molnár[1], Yongqing Li[3], Jianhua Zhao[2]*, and Peng Xiong[1]*

[1]*Department of Physics, Florida State University, Tallahassee, Florida 32306, USA*

[2]*State Key Laboratory of Superlattices and Microstructures, Institute of Semiconductors, Chinese Academy of Sciences, Beijing, 100083, China*

[3]*Beijing National Laboratory for Condensed Matter Physics, Institute of Physics, Chinese Academy of Sciences, Beijing, 100190, China*

[4]*College of Applied Sciences, Beijing University of Technology, Beijing 100124, China*

*\*Email: jhzhao@red.semi.ac.cn, pxiong@fsu.edu*

# Contents

## Supplementary Figure 1: Characterizations of the (Ga,Mn)As thin films

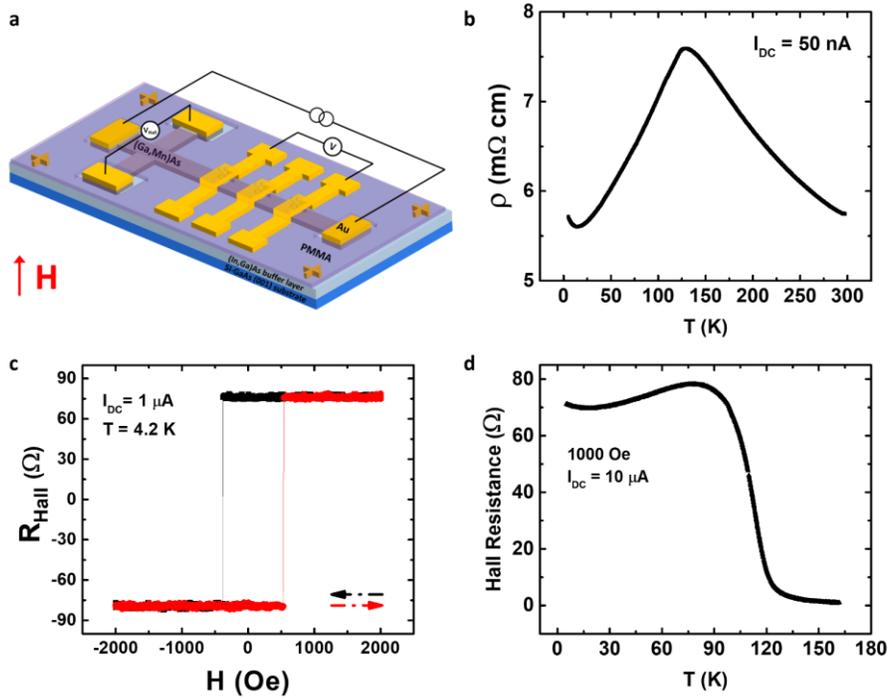

**Fig. S1 | Characterization of the (Ga,Mn)As thin films. a**, Schematics of the experimental setup for magnetoresistance and Hall measurements of the (Ga,Mn)As films. **b**, Resistivity versus temperature for a (Ga,Mn)As thin film. **c**, Hall measurements of (Ga,Mn)As with perpendicular magnetic field. **d**, Hall resistance versus temperature at 1000 Oe.

The (Ga,Mn)As thin films were characterized with magnetoresistance and Hall effect measurements, as shown in the schematics in Fig. S1a. The R(T) behavior of the (Ga,Mn)As can be used to infer the Curie temperature[1]. As shown in Fig. S1b, the (Ga,Mn)As has a Curie temperature about 125 K. The Hall resistance in Fig. S1c evidences the perpendicular magnetic anisotropy of the film, and shows a coercive field of 460 Oe and Hall resistance of 80 Ω. Fig. S1d shows the variation of Hall resistance with temperature, indicating a Curie temperature consistent with that from the R(T).

## Supplementary Figure 2: Stability of the junctions

The magnetoconductance of a (Ga,Mn)As/AHPA-L molecule/Au junction was measured right after device fabrication (within a day). The result is shown in Fig. S2a. The same device, after being stored in a desiccator at room temperature for four months, was measured again under essentially the same conditions. The result is shown in Fig. S2b. The similar results demonstrate remarkable stability of the molecular junctions in ambient (dry) environment.

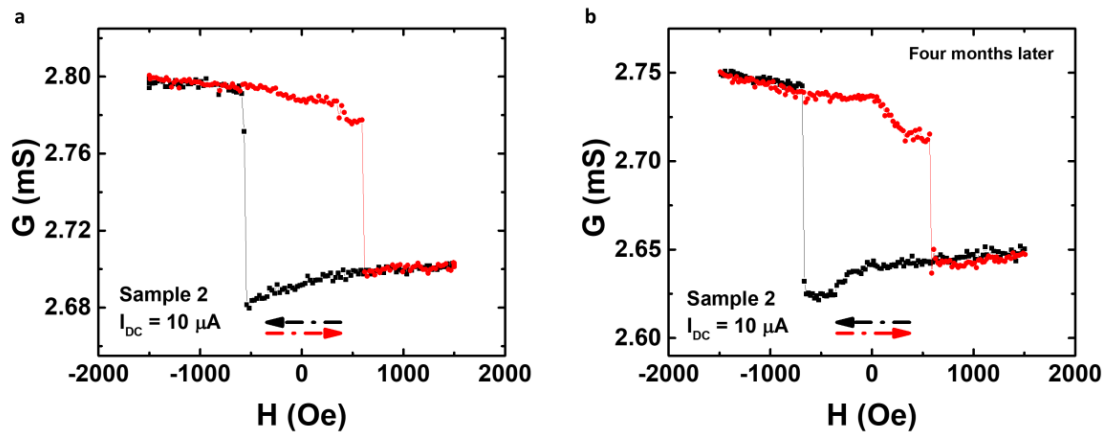

**Fig. S2 | The stability of the junctions.** Junction magnetoconductance of a device measured with a DC bias of 10 µA under the same conditions **a**, right after device fabrication and **b**, four months later.

## Supplementary Figure 3: Bias dependences of $\Delta G_J$ in sample 3

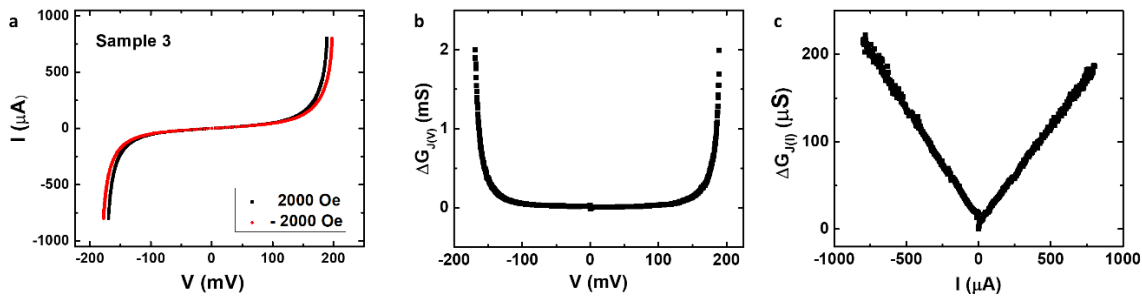

**Fig. S3 | Bias dependence of junction magnetoconductance in sample 3. a**, I-V characteristics of the junction in perpendicular magnetic fields of ±2000 Oe. $\Delta G_J$ as a function of bias voltage (**b**) and bias current (**c**). The black squares are extracted with the same way as in sample 1.

The I-V characteristics and the bias dependences of sample 3 are shown in Fig. S3. They exhibit similar behaviors as those of sample 1 shown in main text.

## Supplementary Figure 4: Minimal asymmetry of I-V curves of the molecular junctions

In Fig. S4, we replot the I-V curves from Fig. 2c in the main text, with the negative bias portion inverted, in order to show the very small offset between the positive and negative bias regions of the I-V curves for both saturation fields. The minimal asymmetry in the I-V is consistent with the argument in Note2 of a very thin Schottky barrier and dominant direct tunneling in the junction.

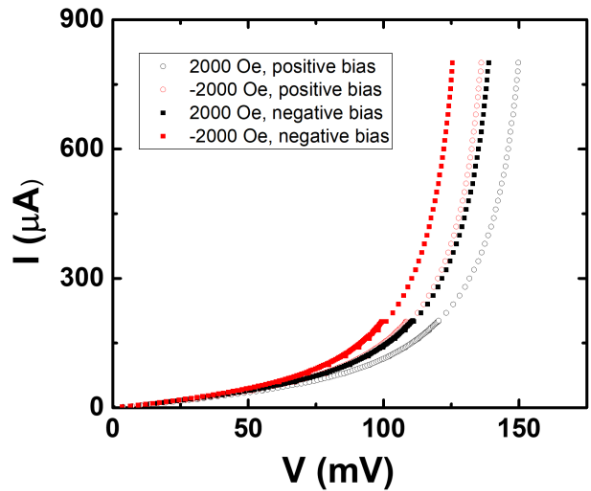

**Fig. S4 | The minimal asymmetry of I-V curves of the junction.** The I-V curves from Fig. 2c, with the negative bias portion inverted for a direct comparison with the positive bias portion.

## Supplementary Figure 5: Background subtraction for MC data

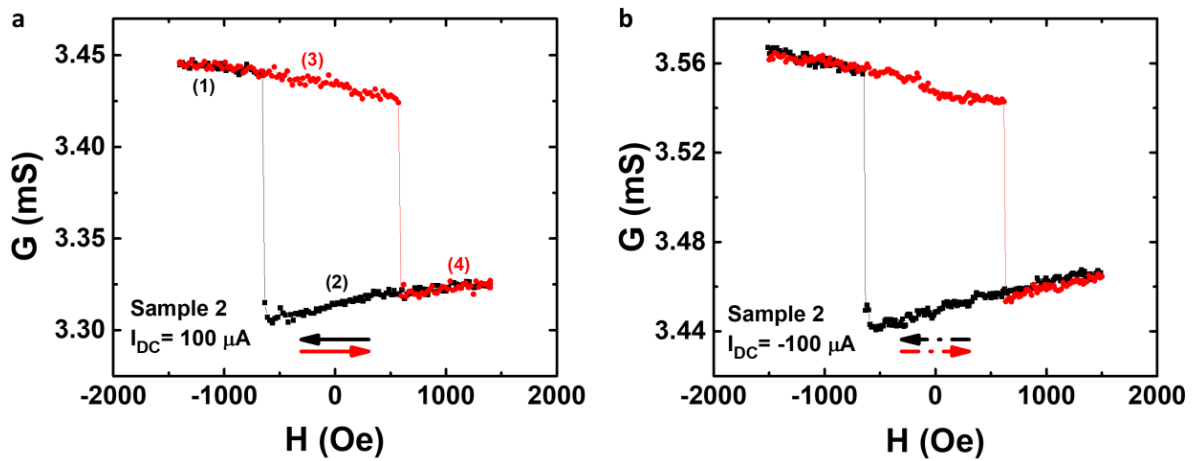

**Fig. S5 | The raw MC data for Fig. 4e in the main text. a** and **b** correspond to DC bias of 100 μA and -100 μA respectively.

In some junctions (e.g., sample 2 in the main text), the raw MC data contain a symmetric hysteretic component. To obtain the antisymmetric component of the signal, a data set is divided into four separate sections and each fitted to a proper polynomial ($r^2$ greater than 0.9). The antisymmetric component is then obtained by subtracting the background without the intercept. The resulting 'net' signals for this device are shown in Fig. 4e in the main text.

## Supplementary Note 1: Estimation of the molecular coverage in the junctions

The molecular coverage of the junctions is estimated via analysis of the junction conductance, based on a model of parallel conduction through areas with and without AHPA-L molecule SAM in a junction. The specific junction conductance is obtained: $g_J = \frac{G_J}{area}$. By the definition of parallel conduction, we have $g_J = C_p \cdot g_m + (1 - C_p)g_{Au}$, where $C_p$ is the coverage of the AHPA-L molecules in the junction, $g_m$ is the specific junction conductance with full coverage of the molecules, $g_{Au}$ is the specific junction conductance of direct contact between Au and (Ga,Mn)As without any molecules, and $g_J$ is the actual specific junction conductance. To estimate the lower limit of $C_p$, we have $(1 - C_p)g_{Au} < g_J$, thus $C_p > 1 - \frac{g_J}{g_{Au}} = 1 - \frac{0.53}{60} = 99\%$ based on the data from Fig. 2e for the control sample and MC with 1 µA as the bias current in sample 1.

## Supplementary Note 2: Estimation of the Schottky barrier between Au and (Ga,Mn)As

Since the (Ga,Mn)As is highly doped, the depletion width of the Schottky barrier is expected to be very thin. The depletion width can be estimated as follows:

$$x_d = \sqrt{\frac{2\varepsilon\phi}{qN_D}} = \sqrt{\frac{2 \times 12.9 \times 8.8 \times 10^{-14} \times 0.42}{1.6 \times 10^{-19} \times 10^{21}}}\ cm = 0.77\ nm.$$

Here $\varepsilon$ is the dielectric constant $\varepsilon = \varepsilon_r \varepsilon_0$. $\phi$ is the built-in potential and is taken as 0.42 V for the p-GaAs and Au contact. $q$ is the electronic charge. $N_D$ is the doping level of (Ga,Mn)As, which is very high in this case. The very small depletion width implies that direct tunneling dominates the transport between the two contacts, where the I-V curves should be mostly linear. Thus the nonlinear I-V curves shown in Fig. 2c are consistent with the significant molecular coverage in the junctions calculated above in Note 1.

## Supplementary Note 3: Theoretical derivations of the junction conductance $G_J$ and conductance change $\Delta G_J$

Based on the theoretical model of Dalum and Hedegård for a two-terminal CISS device[2], we can derive the bias dependences of the junction conductance and magnetoconductance from Eq. 1 in the main text. Let $\Delta\mu = \mu_R - \mu_L = eV$, where $V$ is the bias voltage applied across the junction, and (based on the I-V characteristic of the junction in Fig. 2c) expand the Fermi distribution function to the first three orders, we have

$$n_F(E - \mu_L) - n_F(E - \mu_R) = n_F(E - \Delta\mu) - n_F(E) = \alpha V + \beta V^2 + \gamma V^3, \tag{S1}$$

where $\alpha = \dfrac{e^{\frac{E}{k_B T}}}{\frac{k_B T}{e}\left(1+e^{\frac{E}{k_B T}}\right)^2}$, $\beta = \dfrac{e^{\frac{E}{k_B T}}\left(-1+e^{\frac{E}{k_B T}}\right)}{\frac{2k_B T^2}{e^2}\left(1+e^{\frac{E}{k_B T}}\right)^3}$, $\gamma = \dfrac{e^{\frac{E}{k_B T}}\left(1-4e^{\frac{E}{k_B T}}+e^{\frac{2E}{k_B T}}\right)}{\frac{6k_B T^3}{e^3}\left(1+e^{\frac{E}{k_B T}}\right)^4}$.

In the integral, both $T_{LR}^0$ and $A$ are energy $E$ dependent. Without explicit expression, we write

$$I_{\uparrow\downarrow} = \mathbf{V}\int \alpha T_{\uparrow\downarrow} \frac{dE}{2\pi} + \mathbf{V}^2 \int \beta T_{\uparrow\downarrow} \frac{dE}{2\pi} + \mathbf{V}^3 \int \gamma T_{\uparrow\downarrow} \frac{dE}{2\pi}. \tag{S2}$$

$$\Delta I = I_\downarrow - I_\uparrow = \mathbf{V}\int \alpha \Delta T \frac{dE}{2\pi} + \mathbf{V}^2 \int \beta \Delta T \frac{dE}{2\pi} + \mathbf{V}^3 \int \gamma \Delta T \frac{dE}{2\pi}. \tag{S3}$$

Here, $T_{\uparrow\downarrow} = T_{LR}^0(1 \pm A\vec{m} \cdot \vec{a}_{\uparrow\downarrow})$ and $\Delta T = T_\uparrow - T_\downarrow = T_{LR}^0 A\vec{m} \cdot (\vec{a}_\uparrow + \vec{a}_\downarrow)$. For opposite magnetizations of the (Ga,Mn)As, the junction conductance has two distinct states:

$$G_{\uparrow\downarrow} = \frac{I_{\uparrow\downarrow}}{V} = \int \alpha T_{\uparrow\downarrow} \frac{dE}{2\pi} + \mathbf{V}\int \beta T_{\uparrow\downarrow} \frac{dE}{2\pi} + \mathbf{V}^2 \int \gamma T_{\uparrow\downarrow} \frac{dE}{2\pi}. \tag{S4}$$

$$\Delta G = \Delta G_{J(V)} = \frac{I_\downarrow - I_\uparrow}{V} = \int \alpha \Delta T \frac{dE}{2\pi} + V \int \beta \Delta T \frac{dE}{2\pi} + V^2 \int \gamma \Delta T \frac{dE}{2\pi}. \tag{S5}$$

We emphasize that, $G_{\uparrow\downarrow}$ in Eq. S4 is not the same as $G_J$: $G_{\uparrow\downarrow}$ denote the conductances only through the chiral molecules, while $G_J$ is the overall junction conductance including that of the direct contact between Au and (Ga,Mn)As.